\documentclass[10pt,twocolumn,letterpaper]{article}

\usepackage{iccv}              
\usepackage{bm}
\usepackage{graphicx}
\usepackage{float}
\usepackage{subfloat}
\usepackage{amssymb}
\usepackage{amsmath}
\usepackage{amsthm}
\usepackage{booktabs}
\usepackage[switch]{lineno}
\usepackage{xcolor}
\usepackage{multirow}
\usepackage{bm}
\usepackage{bbm}
\usepackage{dsfont}

\definecolor{iccvblue}{rgb}{0.21,0.49,0.74}
\usepackage[pagebackref,breaklinks,colorlinks,allcolors=iccvblue]{hyperref}

\title{Learnable Retrieval Enhanced Visual-Text Alignment and Fusion \\ for Radiology Report Generation}

\author{
Qin Zhou\textsuperscript{\rm 1,2}\thanks{These authors contributed equally.}, 
Guoyan Liang\textsuperscript{\rm 3,4}\footnotemark[1], 
Xindi Li\textsuperscript{\rm 3,4}, 
Jingyuan Chen\textsuperscript{\rm 3}, 
Wang Zhe\textsuperscript{\rm 1,2}\thanks{Corresponding Authors.}, 
Chang Yao\textsuperscript{\rm 3,4}\footnotemark[2], 
Sai Wu\textsuperscript{\rm 3,4}\footnotemark[2]\\
\textsuperscript{\rm 1}Department of Computer Science and Engineering, ECUST, China\\
\textsuperscript{\rm 2}Key Laboratory of Smart Manufacturing in Energy Chemical Process, Ministry of Education, P. R. China \\
\textsuperscript{\rm 3}Zhejiang University, Hangzhou, China\\
\textsuperscript{\rm 4}Hangzhou High-Tech Zone (Binjiang) Institute of Blockchain and Data Security\\
{\tt\small \{sunniezq, wangzhe\}@ecust.edu.cn, \{guoyanl, 12421143, jingyuanchen, changy, wusai\}@zju.edu.cn}
}
\begin{document}
\maketitle

\begin{abstract}
Automated radiology report generation is essential for improving diagnostic efficiency and reducing the workload of medical professionals. However, existing methods face significant challenges, such as disease class imbalance and insufficient cross-modal fusion. To address these issues, we propose the learnable Retrieval Enhanced Visual-Text Alignment and Fusion (REVTAF) framework, which effectively tackles both class imbalance and visual-text fusion in report generation. REVTAF incorporates two core components: (1) a Learnable Retrieval Enhancer (LRE) that utilizes semantic hierarchies from hyperbolic space and intra-batch context through a ranking-based metric. LRE adaptively retrieves the most relevant reference reports, enhancing image representations, particularly for underrepresented (tail) class inputs; and (2) a fine-grained visual-text alignment and fusion strategy that ensures consistency across multi-source cross-attention maps for precise alignment. This component further employs an optimal transport-based cross-attention mechanism to dynamically integrate task-relevant textual knowledge for improved report generation. By combining adaptive retrieval with multi-source alignment and fusion, REVTAF achieves fine-grained visual-text integration under weak image-report level supervision while effectively mitigating data imbalance issues.
The experiments demonstrate that REVTAF outperforms state-of-the-art methods, achieving an average improvement of \textcolor{magenta}{7.4\%} on the MIMIC-CXR dataset and \textcolor{magenta}{2.9\%} on the IU X-Ray dataset. 
Comparisons with mainstream multimodal LLMs (e.g., GPT-series models), further highlight its superiority in radiology report generation\footnote{\url{https://github.com/banbooliang/REVTAF-RRG}}. 
\end{abstract}
\section{Introduction}
Radiology reports are crucial for clinical diagnosis and treatment, but manual generation is time-consuming and heavily dependent on radiologists' expertise, often leading to delays and inconsistencies in medical decision-making. Traditional methods typically view report generation as an extension of image captioning \cite{cornia2020meshed,pan2020x,yu2022coca,zhang2024sam,wang2024multi}. However, unlike image captioning, which assigns equal weight to all visual concepts, radiology reports must prioritize abnormal findings. These findings are often subtle and imbalanced compared to normal details, causing models to overlook critical diagnostic cues \cite{irvin2019chexpert}. Furthermore, radiologists typically generate reports sentence-by-sentence, each describing an imaging pattern based on regional abnormal findings, which is not originally provided in the training data. The lack of detailed region-sentence correspondence in the training set, leading to insufficient visual-text alignment and fusion during report generation. 
Recent methods have attempted to address these challenges by incorporating external knowledge through auxiliary tasks like classification or by integrating reference reports via retrieval-based algorithms \cite{liu2021exploring,jin2024promptmrg,li2023unify}. 
However, these strategies often fail to provide precise reference information and neglect fine-grained visual-text alignment, potentially introducing irrelevant or incorrect knowledge and further compromising the quality of visual-text integration.
To address these issues, we propose the learnable Retrieval Enhanced Visual-Text Alignment and Fusion (REVTAF) framework. Specifically, we introduce a Learnable Retrieval Enhancer (LRE) module that dynamically retrieves the most relevant reference reports for each input image, acting as Global Reference Prompts (GRPs). This enhances the visual representation, particularly for tail classes, which require precise reference information. In LRE, we map visual features into the hyperbolic space to utilize the semantic hierarchy of images and develop a ranking-based metric to explore intra-batch contextual relationships. We use the hashing distance between multi-class disease labels as a strong supervisory signal to guide the metric learning process.


To improve cross-modal information integration under weak image-report level supervision, we propose a novel Fine-grained Visual-Text Alignment and Fusion (FVTAF) module. This module introduces a Fine-grained Cross-modal Consistency (FCC) constraint that aligns visual and text data using semantic similarities between multi-source text prompts (i.e., GRPs and LRPs). The report-level Global Reference Prompts (GRPs) are generated by the above-mentioned LRE module, while entity-specific Local Reference Prompts (LRPs) are derived from the MedKLIP foundation model \cite{wu2023medklipmedicalknowledgeenhanced}. Additionally, we incorporate the optimal transport-based cross-attention mechanism to extract relevant information from the multi-source text prompts, ensuring thorough feature fusion.

Our main contributions are outlined as follows:
\begin{itemize}
    \item We propose a novel framework that combines a Learnable Retrieval Enhancer and a Fine-grained Visual-Text Alignment and Fusion module to simultaneously tackle the challenges of class imbalance and insufficient cross-modal fusion.
    \item We pioneer a learnable solution for adaptively retrieving the most relevant reference report for each input image, effectively augmenting the representation of visual input, particularly for tail classes.
    \item We design a novel visual-text alignment and fusion module that integrates Fine-grained Cross-modal Consistency with an optimized cross-attention mechanism, enabling more effective visual-text integration.
    \item Extensive comparisons with state-of-the-art radiology report generation methods and multimodal LLMs, as well as ablative studies, consistently demonstrate the superior performance of our approach.
\end{itemize}

\begin{figure*}[bht]
\centering
\includegraphics[width=0.82\textwidth]{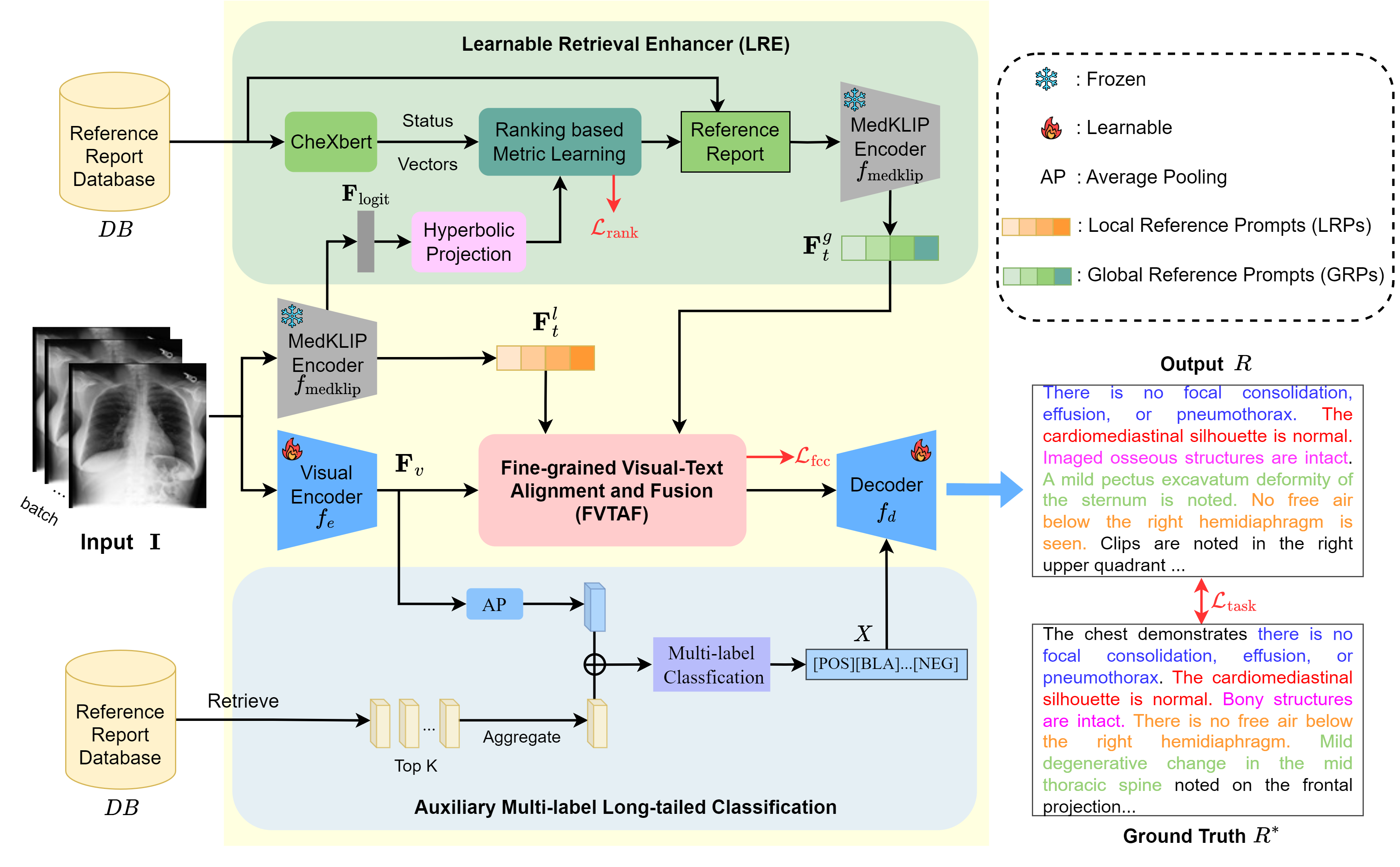}
\caption{Overview of the proposed Retrieval Enhanced Visual-Text Alignment and Fusion (REVTAF) framework.}
\label{fig1:env}
\end{figure*}

\section{Related Work}
\subsection{Image Captioning}
Generating radiology reports shares many similarities with the image captioning task, which focuses on producing concise textual descriptions of images, and has garnered significant research interest in recent years \cite{lu2017knowing,anderson2018bottom,pan2020x,wang2023metransformer}.
Most approaches in image caption rely on an encoder-decoder framework: an image encoder first extracts visual features, which are then passed to a text decoder to generate the final captions.
Early studies \cite{lu2017knowing,anderson2018bottom,li2023zero} primarily employed Long Short-Term Memory (LSTM) networks \cite{graves2012long} and Convolutional Neural Networks (CNN) \cite{o2015introduction} to tackle this task. 
Recently, transformer models that leverage attention mechanisms \cite{vaswani2017attention} have become prevalent due to their superior capability in processing intricate vision and language features \cite{cornia2020meshed,pan2020x,wang2023metransformer}.
Additionally, innovative training strategies such as reinforcement learning \cite{rennie2017self} and adversarial training \cite{dai2017towards} have further boosted performance in image captioning. 
Recent advancements in foundation models have highlighted the effectiveness of large-scale visual-language pre-training \cite{hu2022scaling,yu2022coca,wang2022medclipcontrastivelearningunpaired} for image captioning tasks. However, these methods often fail to incorporate domain-specific medical knowledge and lack region-to-sentence level correspondence, which significantly limits their applicability to radiology report generation \cite{wang2023metransformer}.

\subsection{Radiology Report Generation}
Radiology reports typically comprise multiple descriptive sentences that detail the findings observed in radiological images. Unlike standard image captioning tasks, radiology report generation requires not only producing longer outputs but also achieving greater precision in describing region-specific imaging findings. To tackle these challenges, various methods have been proposed \cite{chen2020generating,yan2022clinical,li2023dynamic,huang2023kiut,jin2024promptmrg}.

Some works leverage memory-based approaches to capture and retain critical information. For instance, R2Gen \cite{chen2020generating} and R2GenCMN \cite{chen2022cross} enhance the standard encoder-decoder framework by separately fusing image and caption data using LSTM and cross-modal memory networks (CMN). 
These models utilize shared memory to record the alignment between images and text, thereby facilitating effective cross-modal interactions. 
Inspired by this idea, XproNet \cite{wang2022cross} employs a shared cross-modal prototype matrix that serves as external knowledge, capturing and embedding cross-modal prototypes to improve report generation.

Other approaches integrate additional knowledge sources to assist the generation process. 
Clinical-BERT \cite{yan2022clinical} introduces a visual-language pre-trained model that incorporates medical domain knowledge to boost performance. 
Similarly, \cite{liu2021exploring} explores the integration of both posterior and prior knowledge distilled from visual cues, medical graphs, and retrieved reports. 
PromptMRG \cite{jin2024promptmrg} incorporates auxiliary multi-disease classification task to improve diagnostic accuracy. 
Some approaches focus on modeling relational context across text or key visual information to enhance the report generation.
The DCL model \cite{li2023dynamic} fuses information from a pre-constructed knowledge graph that encodes relationships between caption words, while KiUT \cite{huang2023kiut} introduces a knowledge-injected U-transformer that learns multi-level visual representations and adaptively distills contextual and clinical knowledge for precise word prediction.
EKAGen \cite{bu2024instance} converts expert reports into an embedding space to prevent the loss of salient features and to enhance focus on key regions. 
Despite progress, current methods still face challenges in \textit{balancing disease categories and effectively integrating visual-text data, limiting radiology report quality.}

\section{Method}
This section delves into the specifics of our learnable Retrieval Enhanced Visual-Text Alignment and Fusion (REVTAF) framework. Initially, we will introduce the essential background on radiology report generation to provide context for our approach.
\subsection{Preliminaries}
Radiology report generation refers to automatically generating structured and coherent radiology reports from medical imaging data.
Formally, given a 2D radiology image $\mathbf{I}$, the model is tasked with interpreting the image and generating a descriptive radiology report $R = \{r_1, r_2, ..., r_{T} \}$, where $T$ is the length of the report, and $\mathbb{V}$ represents the vocabulary, each $r_t \in \mathbb{V}$ is a token.
The entire recursive generation process is formulated as follows,
\begin{equation}
    p(R | \mathbf{I}) = \prod_{t=1} p(r_{t} | r_1, ..., r_{t-1}, \mathbf{I}).
\end{equation}

The report generation process is optimized by minimizing the cross-entropy loss,

\begin{equation}
    \mathcal{L}_{\text{task}} = - \sum \limits_{t=1}^T log(p(r_t | r_{1:t-1})).
\end{equation}

Existing approaches \cite{chen2023fine,jin2024promptmrg} for radiology report generation use pre-trained models (e.g., CLIP) to compute similarity scores between an image and training reports, retrieving the most relevant report to guide the generation process. 
However, these methods face several challenges. First, the absence of domain-specific knowledge in pre-trained models often leads to suboptimal reference selection. Second, imbalanced disease labels result in biased reports that neglect rare diseases and subtle anomalies. Additionally, the lack of fine-grained alignment hinders effective visual-text feature fusion under weak image-report level supervision. Our proposed REVTAF framework effectively addresses the aforementioned challenges, establishing a new benchmark for the community.


\subsection{Framework Overview}
Figure \ref{fig1:env} outlines the workflow of our REVTAF framework. Given an input image \(\mathbf{I} \in \mathbb{R}^{H \times W}\) and a reference report database \(DB = \{R_1, \dots, R_{N_D}\}\), where $N_D$ is the number of reference reports in the database, REVTAF employs an encoder-decoder architecture for radiology report generation. The visual encoder \(f_e\) extracts intermediate features \(\mathbf{F}_v \in \mathbb{R}^{H_v \times W_v \times d_v} = f_e(\mathbf{I})\), while the text decoder \(f_d\) generates the final report.  
To enhance cross-modal alignment, we introduce two components: 

(1) entity-specific Local Reference Prompts (LRPs): We adopt the pre-trained MedKLIP model to produce local text prompts \(\mathbf{F}_t^l \in \mathbb{R}^{M \times d_l}\), where \(M\) corresponds to refined entity classes as described in \cite{yu2022anatomy}.

(2) report-level Global Reference Prompts (GRPs): A Learnable Retrieval Enhancer (LRE) adaptively retrieves the most relevant report from \(DB\) by learning a hyperbolic space ranking metric that captures image semantic hierarchies and intra-batch context. Denote $R_{ref}$ as the retrieved most relevant report. Then we generate the corresponding report-level GRPs as \(\mathbf{F}_t^g \in \mathbb{R}^{N \times d_g} = f_{\text{medklip}}(R_{ref})\), with \(N\) denoting the maximum sentence count in \(DB\), $f_{\text{medklip}}$ representing the pretrained MedKLIP encoder.

The Fine-grained Visual-Text Alignment and Fusion (FVTAF) module integrates multi-source prompts (\(\mathbf{F}_t^l\) and \(\mathbf{F}_t^g\)) with visual features \(\mathbf{F}_v\). It introduces a novel Fine-grained Cross-modal Consistency constraint for visual-text alignment, alongside an optimal transport-based cross-attention mechanism to fuse \(\{\mathbf{F}_v, \mathbf{F}_t^g\}\) into global cross-modal features \(\mathbf{F}_c^g\). Similarly, local cross-modal features \(\mathbf{F}_c^l\) are generated from \(\{\mathbf{F}_v, \mathbf{F}_t^l\}\). 
Following \cite{jin2024promptmrg}, an auxiliary multi-label long-tailed classification branch produces disease-related prompts $X$ to mitigate data imbalance. Finally, the decoder \(f_d\) synthesizes the fused visual-text features \( \mathbf{F}_c = \oplus(\mathbf{F}_c^g,\mathbf{F}_c^l)\) ( where $\oplus(\cdot)$ denotes concatenation along the feature dimension) and \(X\) to generate the final report \(R = f_d(\mathbf{F}_c, X)\). Details of the proposed LRE and FVTAF modules are discussed in subsequent sections.

\subsection{Learnable Retrieval Enhancer}
The relevance of the retrieved reference report is crucial for effective guidance, particularly in cases of long-tailed distribution \cite{jin2024promptmrg}. Current methods \cite{chen2023fine,jin2024promptmrg} rely on reports linked to database images with similar pooled CLIP features for reference. However, this approach lacks medical domain knowledge and overlooks important spatial details. To address these issues, we propose the Learnable Retrieval Enhancer (LRE) to enhance retrieval relevance. 

First, we leverage the medical foundation model MedKLIP to extract visual features from training images and text features from reports. MedKLIP generates entity-specific logits \(\mathbf{F}_{\text{logit}} \in \mathbb{R}^{B \times M}\) (where \(B\) = batch size, \(M = 75\) entity classes), highlighting abnormal findings in input images. 
Building on this, we design a hyperbolic space ranking-based metric to learn hierarchical visual similarities among training images within current mini-batch, supervised by hashing distances derived from disease labels related to the corresponding reports.  

\textbf{Hyperbolic Space Distance.} 
Given the remarkable structural consistency of medical images, where each organ exhibits a well-defined spatial arrangement, we project entity-specific logits \(\mathbf{F}_{\text{logit}}\) into hyperbolic space using a Hyperbolic Neural Network (HNN) \cite{yang2024improving}. HNN can adaptively learn hierarchical representations tailored to anatomical organization. Let the resulting hyperbolic features be denoted as \(\mathbf{H} \in \mathbb{R}^{B \times d_h} = \text{HNN}(\mathbf{F}_{\text{logit}})\), where \(d_h\) is the hyperbolic embedding dimension.  
In the Poincaré ball model with curvature $c$ (\(\mathbb{B}^c\)), the geodesic distance \(d_{\mathbb{B}}^c(\mathbf{x}, \mathbf{y})\) between two points \(\mathbf{x}, \mathbf{y} \in \mathbb{B}^c\) is computed as:  
\begin{equation}
d_{\mathbb{B}}^c (\mathbf{x}, \mathbf{y}) = \frac{2}{\sqrt{c}} \tanh^{-1} \left( \sqrt{c} \, \| \mathbf{x} \oplus_c \mathbf{y} \| \right),
\label{eq:hnn_dist}
\end{equation} 
where \(\oplus_c\) denotes the Möbius addition operation:  
\begin{equation}
\mathbf{x} \oplus_c \mathbf{y} = \frac{(1 + 2c\mathbf{x}^\top\mathbf{y} + c\|\mathbf{y}\|^2)\mathbf{x} + (1 - c\|\mathbf{x}\|^2)\mathbf{y}}{1 + 2c\mathbf{x}^\top\mathbf{y} + c^2\|\mathbf{x}\|^2\|\mathbf{y}\|^2}.
\end{equation}

Using Eq.~(\ref{eq:hnn_dist}), we compute the pairwise distance matrix $\hat{D}  \in \mathbb{R}^{B \times B} = \{\hat{D}_{ij}, i,j \in 1,\cdots, B\}$ for the entire batch, where each element $\hat{D}_{ij}$ represents the hyperbolic distance between samples $i$ and $j$. Let $\mathbf{h}_i$ and $\mathbf{h}_j$ denote the hyperbolic features of the $i$-th and $j$-th samples in current batch. Then $\hat{D}_{ij}$ is computed as,
\begin{equation}
    \hat{D}_{ij} = d_{\mathbb{B}}^c (\mathbf{h}_i, \mathbf{h}_j).
\end{equation}

To incorporate semantic guidance into hyperbolic feature learning, we leverage the semantic similarity between paired radiology reports as supervisory signal. Specifically, for each sample, we extract structured classification labels from its corresponding report \(R\) using CheXbert \cite{smit2020chexbert}, which maps $R$ to \(K = 18\) predefined disease categories. Each category is assigned one of four statuses: Blank (unmentioned), Positive (disease present), Negative (disease absent), or Uncertain. This results in a 72-dimensional status vector per report, with each element in $\{0,1\}$. Then we can calculate the ground-truth semantic distance matrix $D \in \mathbb{R}^{B \times B}$ using hashing distance between their status vectors. 
Let \(\mathbf{v}_i, \mathbf{v}_j \in \{0,1\}^{72}\) denote the status vectors for the \(i\)-th and \(j\)-th samples, respectively, their semantic distance $D_{ij}$ can be calculated as, 
\begin{equation}
    D_{ij} = \sum \limits^{72}_{k=1} \mathds{1}[\mathbf{v}_{i,k} \neq \mathbf{v}_{j,k}],
\end{equation}
where $\mathds{1}[\cdot]$ is the indicator function that evaluates to 1 when the inside condition is true, and 0 otherwise. $\mathbf{v}_{i,k},\mathbf{v}_{j,k}$ denote the $k$-th values in $\mathbf{v}_{i}$ and $\mathbf{v}_{j}$, respectively.

\textbf{Ranking based Metric Learning.}
While a naive approach would compute the Euclidean distance between the hyperbolic distance matrix \(\hat{D}\) and the ground-truth pairwise semantic distance matrix \(D\), this method ignores contextual relationships across samples and is prone to outlier sensitivity. To address this, we adopt ranking-based metric learning as follows:  
   For each row of the ground-truth distance matrix \(D\), we sort the entries in ascending order. Denote \(\pi_{i}\) as the index with the smallest hamming distance relative to the \(i\)-th sample within the batch, we then compute a cross-entropy loss by treating the hyperbolic distance matrix \(\hat{D}\) as predictions and the ground-truth index vector \(\Pi = \{\pi_{i}\}\) as targets: 
   \begin{equation}
   \mathcal{L}_{\text{rank}} = \text{CE}(\hat{D}, \Pi),
   \end{equation}  
   where \(\text{CE}(\cdot)\) denotes the CrossEntropy loss, penalizing deviations between the predicted hyperbolic distance rankings and the semantic similarity rankings. Note that diagonal values are omitted before ranking to prevent self-comparison.
During training, ground-truth hashing distances between reports are used to select the most relevant report as the Global Reference Prompts (GRPs). During inference, the report related to the most similar database image in the learned hyperbolic space is retrieved, generating the report-level GRPs. GRPs then serve as a text input for the subsequent alignment and fusion module.
   


\begin{figure}
\centering
\includegraphics[width=0.42\textwidth]{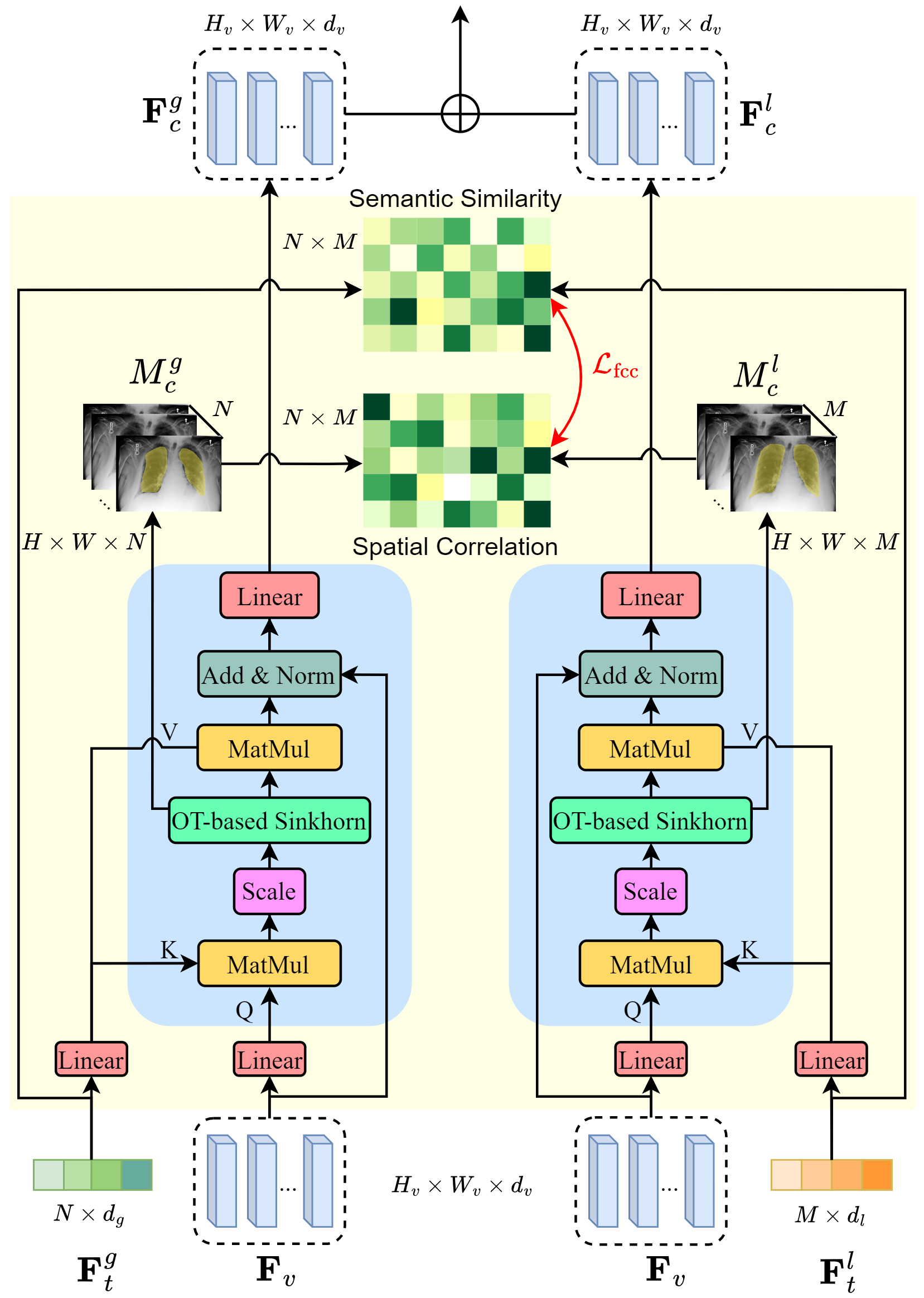}
\caption{Illustration of the Fine-grained Visual-Text Alignment and Fusion (FVTAF) module.}
\label{fig2:env}
\end{figure}
\subsection{Fine-grained Visual-Text Alignment and Fusion}
The image-report level weak supervision in medical report generation hinders fine-grained region-sentence alignment. To address this, we propose the Fine-grained Visual-Text Alignment and Fusion (FVTAF) module, which enhances visual input by integrating knowledge from multi-source text inputs, as shown in Figure \ref{fig2:env}. In addition to the Global Reference Prompts (GRPs) generated above to capture report-level information from the most similar reference image, we introduce entity-specific Local Reference Prompts (LRPs). These LRPs describe disease findings tied to regional imaging patterns within the input image. Specifically, we leverage the MedKLIP foundation model to generate triplets encoding the position and existence of \( M = 75 \) predefined entities.  
Formally, each report is represented as a set of triplets:  
\begin{equation}
R = \{(entity_m, position_m, exist_m)\}_{m=1}^M,
\end{equation} 
where each triplet for the \( m \)-th entity is concatenated and fed into the MedKLIP textual encoder to produce the corresponding local reference prompt. The overall entity-specific LRPs are constructed by concatenating textual embeddings from all entities:  
$
\mathbf{F}^{l}_{t} = \left[ \mathbf{F}^{l}_{t}(1), \mathbf{F}^{l}_{t}(2), \dots, \mathbf{F}^{l}_{t}(M) \right],
$
where each \( \mathbf{F}^{l}_{t}(m) \) is computed as:  
\begin{equation}
\mathbf{F}^{l}_{t}(m) = f_{\text{medklip}}  \big([entity_m, position_m, exist_m]\big).
\end{equation}
\textbf{Multi-source Cross-modal Fusion.} Given the visual features $\mathbf{F}_{v}$, report-level Global Reference Prompts (GRPs) $\mathbf{F}^{g}_{t}$ and entity-specific Local Reference Prompts (LRPs) $\mathbf{F}^{l}_{t}$, we introduce two visual-text cross-attention branches to enhance the query visual features with knowledge from both GRPs and LRPs. During the visual-text cross attention modeling, instead of adopting the traditional cross-attention mechanism \cite{liu2024groundingdinomarryingdino,wang2024multi}, we resort to \cite{ye2024otseg} to utilize the Multi-Prompts Sinkhorn Attention (MPSA) to better incorporate textual knowledge more relevant to the visual embeddings. The MPSA mechanism leverages optimal transport theory to reweight query-key cross-attention with global context, effectively extracting relevant knowledge and filtering out unrelated textual information to better adapt to the input image. For details of the MPSA mechanism, please refer to \cite{ye2024otseg}. Formally, the fused global visual-text features and cross-attention maps can be obtained as,
\begin{equation}
\begin{aligned}
    & \mathbf{F}_{c}^{g}, M_c^{g} = MPSA(\mathbf{Q}\mathbf{K}^T) \mathbf{V}, \\
      & \mathbf{Q} = \phi_q (\mathbf{F}_{v}), \mathbf{K} = \phi_k (\mathbf{F}_{t}^{g}), \mathbf{V} = \phi_v (\mathbf{F}_{t}^{g}),
     \label{eq:msa}
\end{aligned}
\end{equation}
Similarly, we can generate the fused local viusal-text features and attention maps as $\mathbf{F}_{c}^{l}, M_c^{l}$.

To ensure the generated report remains faithful to the input image, we introduce residual connections to retain the original visual features in the fused representations. Specifically, for both global (\(\mathbf{F}_{c}^{g}\)) and local (\(\mathbf{F}_{c}^{l}\)) cross-modal features, the final fused features are computed as:  
\begin{equation}
\begin{aligned}
  \mathbf{F}_{c}^g &= f_{proj}\left(\text{LayerNorm}\left(\mathbf{F}_{c}^{g} + \mathbf{F}_v\right)\right), \\
  \mathbf{F}_{c}^l &= f_{proj}\left(\text{LayerNorm}\left(\mathbf{F}_{c}^{l} + \mathbf{F}_v\right)\right),
\end{aligned}
\end{equation} 
where \(\text{LayerNorm}(\cdot)\) denotes layer normalization to stabilize training, and \(f_{proj}\) is a learnable linear projection layer.  
The global and local fused features \(\mathbf{F}_{c}^g\) and \(\mathbf{F}_{c}^l\) are then concatenated along the feature dimension to form the final fused representation   
$\mathbf{F}_{c}$, which is passed to the transformer decoder for report generation.  

\begin{table*}[htb]
    \centering
    \scalebox{0.8}{
    \begin{tabular}{l|l|cccccc|ccc|c}
        \toprule
        \multirow{2}{*}{Model} & \multirow{2}{*}{Year}&\multicolumn{6}{c|}{NLG Metrics}  & \multicolumn{3}{c|}{CE Metrics }&\multirow{2}{*}{Avg}   \\
        \cline{ 3-11 }
                & & BLEU-1 &BLEU-2 &BLEU-3 &BLEU-4 &METEOR &ROUGE&Precision &Recall &F1   \\
        \midrule
        R2Gen   &ACL 2020       &0.353 &0.218 &0.145 &0.103 &0.142 &0.277 &0.333 &0.273 &0.276&0.236  \\
        M2TR    &ACL 2021       &0.378 &0.232 &0.154 &0.107 &0.145 &0.272 &0.240 &0.428 &0.308&0.252  \\
        MKSG    &MIA 2022      &0.363 &0.228 &0.156	&0.115 &-     &0.284 &0.458 &0.348 &0.371&-  \\
        M2KT     &MIA 2023      &0.386 &0.237 &0.157 &0.111 &-     &0.274 &0.420 &0.339 &0.352 &- \\
        ME       &CVPR 2023      &0.386 &0.250 &0.169 &0.124 &0.152 &0.291 &0.364 &0.309 &0.311&0.262  \\
        KiUT     &CVPR 2023      &0.393 &0.243 &0.159 &0.113 &0.160 &0.285 &0.371 &0.318 &0.321&0.263  \\
        DCL     &CVPR 2023       &-&-&-& 0.109 & 0.150 &0.284 &0.471 & 0.352 & 0.373&-  \\
        UAR      &ICCV 2023      & 0.363 &0.229 &0.158 &0.107 &0.157 &0.289 &-&-&- &-    \\
        HERGen  &ECCV 2024       & 0.395 &0.248 &0.169 &0.122 &0.156 &0.285 &0.415 &0.301 &0.317&0.268 \\
        CVT2Dis.  &Artif.Intell.Med 2022     &0.392 &0.245 &0.169 &0.124 &0.153 &0.285 &0.356 &0.412 &0.384&0.280  \\
         CliBert  &AAAI 2022      &0.383 &0.230 &0.151 &0.106 &0.144 &0.275 &0.397 &0.435 &0.415&0.282  \\
        RGRG     &CVPR 2023      &0.373 &0.249 &0.175 &0.126 &0.168 &0.264 &0.461 & 0.475 & 0.447&0.304  \\
        PromptMRG  &AAAI 2024   &0.398 &0.239 &0.156	&0.112 &0.157 &0.268 &0.501	&0.509 &0.476&0.313   \\
        EKAGen    &CVPR 2024     &0.419 &0.258 &0.170 &0.119 &0.157 &0.287 &0.517 &0.483 &0.499 &0.323 \\
        \midrule
        Ours     &-      &\textbf{0.465} &\textbf{0.318} &\textbf{0.235} &\textbf{0.182} &\textbf{0.199} &\textbf{0.336} &\textbf{0.628}	&\textbf{0.613} &\textbf{0.592}&\textbf{0.397}  \\
        \bottomrule
    \end{tabular}}
    \caption{Comparison with other SOTA methods on the MIMIC-CXR dataset. The best results are highlighted in bold.}
    \label{tab1:env}
\end{table*}
\paragraph{Fine-grained Cross-modal Consistency.}
To mitigate the impact of irrelevant or incorrect descriptions in Global Reference Prompts (GRPs) and Local Reference Prompts (LRPs), we propose the Fine-grained Cross-modal Consistency (FCC) constraint. This constraint is motivated by the assumption that multi-source text prompts sharing similar semantics should produce correlated responses in their cross-modal attention maps (\(M_c^g\) and \(M_c^l\)).   
Given global text features \(\mathbf{F}_t^g\) (from GRPs) and local text features \(\mathbf{F}_t^l\) (from LRPs), we compute a sentence-level semantic similarity matrix \(S \in \mathbb{R}^{N \times M}\). Here, \(S_{nm}\) represents the cosine similarity between the \(n\)-th sentence in GRPs and the \(m\)-th entity in LRPs:  
\begin{equation}
   S_{nm} = \frac{\langle \mathbf{F}^{g}_{t}(n), \mathbf{F}^{l}_{t}(m) \rangle}{\|\mathbf{F}^{g}_{t}(n)\|_2 \|\mathbf{F}^{l}_{t}(m)\|_2},
\end{equation}
where \(\mathbf{F}^{g}_{t}(n)\) and \(\mathbf{F}^{l}_{t}(m)\) denote the \(n\)-th sentence and \(m\)-th entity embeddings in \(\mathbf{F}_t^g\) and \(\mathbf{F}_t^l\), respectively.  
The similarity matrix \(S\) is normalized to the \([0, 1]\) range via a sigmoid function to ensure stable alignment.  

\begin{table*}
    \centering
    \scalebox{0.8}{
    \begin{tabular}{l|l|cccccc|ccc|c}
        \toprule
        \multirow{2}{*}{Model} &\multirow{2}{*}{Year}& \multicolumn{6}{c|}{NLG Metrics}  & \multicolumn{3}{c|}{CE Metrics}&\multirow{2}{*}{Avg}   \\
        \cline{ 3-11 }
        & & BLEU-1 &BLEU-2 &BLEU-3 &BLEU-4 &METEOR &ROUGE&Precision &Recall &F1   \\
        \midrule
        R2Gen   &ACL 2020       &0.289&0.155&0.087&0.052&0.128&0.243&0.151&0.145&0.145&0.155 \\
        M2KT  &MIA 2023         &0.371&0.239&0.151&0.078&0.153&0.261&0.153&0.145&0.145&0.188  \\
        DCL     &CVPR 2023       &0.354&0.230&0.148&0.074&0.152&0.267&0.168&0.167&0.162&0.191  \\
        RGRG    &CVPR 2023     &0.266&0.215&0.147&0.063&0.146&0.180&0.183&0.187&0.180&0.174  \\
        CVT2Dis. &Artif.Intell.Med 2022       &0.383&0.236&0.157&0.082&0.147&0.277&0.174&0.172&0.168&0.200  \\
        PromptMRG &AAAI 2024    &0.401&0.247&\textbf{0.160}&0.098&0.160&0.281&0.213&0.229&0.211&0.222  \\
        \midrule
        Ours    &-       &\textbf{0.420}&\textbf{0.249}&0.159&\textbf{0.107}&\textbf{0.176}&\textbf{0.309}&\textbf{0.286}&\textbf{0.282}&\textbf{0.273}&\textbf{0.251} \\
        \bottomrule
    \end{tabular}}
    \caption{Comparing the performance of our model with other SOTA methods on the IU X-Ray dataset.}
    \label{tab2:env}
\end{table*}
   


We quantify the spatial correlation of cross-attention maps \(M^g_c\) (as calculated in Eq.~\ref{eq:msa}) and \(M^l_c\) using Intersection-over-Union (IoU).   
   For the \(n\)-th attention map \(M^g_c(n)\) in \(M^g_c\) and the \(m\)-th map \(M^l_c(m)\) in \(M^l_c\), their spatial overlap \(O_{nm}\) is computed as:  
   \begin{equation}
   \begin{aligned}
     O_{nm} &= \text{IoU}\left(M^{g}_{c}(n), M^{l}_{c}(m)\right),\\
     &= \frac{\sum \left(M^{g}_{c}(n) \cap M^{l}_{c}(m)\right)}{\sum \left(M^{g}_{c}(n) \cup M^{l}_{c}(m)\right)},
   \end{aligned}
 \end{equation} 
   where \(\cap\) and \(\cup\) denote element-wise minimum and maximum operations, respectively.  
   The FCC loss ($\mathcal{L}_{\text{fcc}}$) penalizes discrepancies between semantic similarity (\(S\)) and spatial correlation (\(O\)):  
   \begin{equation}
   \mathcal{L}_{\text{fcc}} = \frac{1}{N \times M} \sum_{n=1}^N \sum_{m=1}^M \left(1 - S_{nm} \cdot O_{nm}\right),
   \end{equation}
   




The overall training objective $\mathcal{L}$ is the combination of $\mathcal{L}_{\text{task}}$, $\mathcal{L}_{\text{rank}}$, and $\mathcal{L}_{\text{fcc}}$, 
\begin{equation}
    \mathcal{L} = \mathcal{L}_{\text{task}} + \alpha \mathcal{L}_{\text{rank}} + \beta \mathcal{L}_{\text{fcc}},
\end{equation}
where $\alpha$ and $\beta$ are balancing coefficients.

\begin{figure*}[hbt]
\centering
\includegraphics[width=0.80\textwidth]{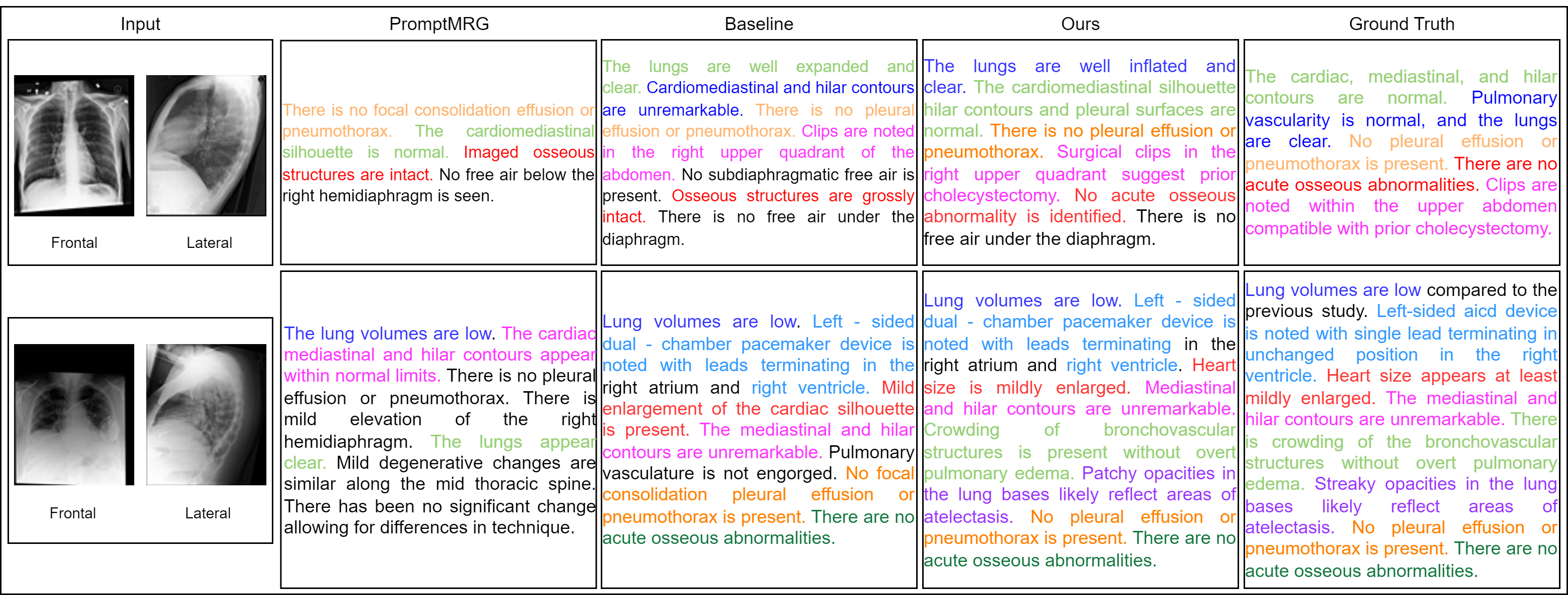}
\caption{Qualitative results on the MIMIC-CXR dataset. Matches with the ground truth are highlighted in the same colors, while inconsistencies are marked in black.}
\label{fig3:env}
\end{figure*}
\begin{table*}[htb]
    \centering
    \scalebox{0.8}{
    \begin{tabular}{c|cc|cccccc|ccc|c}
        \toprule
        \multirow{2}{*}{Setting}&\multirow{2}{*}{LRE}&\multirow{2}{*}{FVTAF}& \multicolumn{6}{c|}{NLG Metrics}  & \multicolumn{3}{c|}{CE Metrics }&\multirow{2}{*}{Avg}   \\
        \cline{ 4-12 }
         &&& BLEU-1 &BLEU-2 &BLEU-3 &BLEU-4 &METEOR &ROUGE&Precision &Recall &F1   \\ 
        \midrule
        Baseline && &0.432&0.283&0.210&0.159&0.177&0.311&0.597&0.586&0.562&0.369 \\
        \midrule
        (a)&  $\checkmark$   & &0.453&0.312&0.224&0.178&0.186&0.325&0.612&0.590&0.571&0.384  \\
        (b)& & $\checkmark$ &     0.457&0.296&0.228&0.175&0.186&0.329&0.607&0.595&0.571&0.383  \\
        \midrule
        (c)& $\checkmark$& $\checkmark$  &\textbf{0.465} &\textbf{0.318} &\textbf{0.235} &\textbf{0.182} &\textbf{0.199} &\textbf{0.336} &\textbf{0.628}	&\textbf{0.613} &\textbf{0.592}&\textbf{0.397}    \\ 
        \bottomrule
    \end{tabular}}
    \caption{Analysis on the effectiveness of each component on MIMIC-CXR test set.}
    \label{tab3:env}
\end{table*}
\begin{table*}
    \centering
    \scalebox{0.81}{
    \begin{tabular}{l|c|cccccc|ccc|c}
        \toprule
        \multirow{2}{*}{LLMs}& \multirow{2}{*}{Time (s)}&\multicolumn{6}{c|}{NLG Metrics}  & \multicolumn{3}{c|}{CE Metrics }&\multirow{2}{*}{Avg}   \\
        \cline{ 3-11 }
         && BLEU-1 &BLEU-2 &BLEU-3 &BLEU-4 &METEOR &ROUGE&Precision &Recall &F1   \\ 
        \midrule
        GPT-4 &15.02&0.286&0.131&0.048&0.00&0.117&0.187&0.293&0.229&0.247&0.192\\
        GPT-4o &8.77&0.326&0.159&0.081&0.047&0.115&0.188&0.146&0.219&0.166&0.161\\
        GPT-4o-mini&7.50&0.254&0.104&0.045&0.022&0.104&0.161&0.063&0.083&0.067&0.100 \\
        GPT-4.5&18.72&0.306&0.140&0.073&0.045&0.112&0.182&0.303&0.224&0.233&0.180
 \\
        \midrule
        Ours&\textbf{3.36}&\textbf{0.379}&\textbf{0.236}&\textbf{0.161}&\textbf{0.119}&\textbf{0.150}&\textbf{0.266}&\textbf{0.466}&\textbf{0.766}&\textbf{0.581}&\textbf{0.347}
 \\
        \bottomrule
    \end{tabular}}
    \caption{Comparison with GPT series multimodal LLMs on randomly sampled MIMIC-CXR test set.}
    \label{tab4:env}
\end{table*}

\section{Experiments}
In this section, we demonstrate the effectiveness of our REVTAF framework through comprehensive comparisons.
Owing to space constraints, we present further ablation experiments in the \textcolor{magenta}{supplementary material}.
\subsection{Experimental Setups}
\paragraph{Datasets and Evaluation Metrics.} 
We evaluate our model on two widely used radiology report generation benchmarks: MIMIC-CXR and IU X-Ray.
The MIMIC-CXR dataset, provided by the Beth Israel Deaconess Medical Center, contains a large-scale collection of chest X-ray images paired with corresponding reports. Following prior studies \cite{chen2022cross,huang2023kiut,jin2024promptmrg}, we split the dataset into 270,790 training samples, 2,130 validation samples, and 3,858 test samples for fair comparison. The IU X-Ray dataset from Indiana University includes 7,470 frontal and lateral chest X-ray images and 3,955 associated reports. Due to the limited positive samples for certain diseases in the test set of the offline split \cite{jin2024promptmrg}, we follow previous methods to evaluate the model trained on MIMIC-CXR, directly on the entire IU X-Ray dataset.

Our model's performance is assessed in two aspects: natural language generation (NLG) and clinical efficacy (CE). For NLG, we measure report quality using BLEU \cite{papineni2002bleu}, METEOR \cite{denkowski2011meteor}, and ROUGE-L \cite{lin2004rouge}. For CE, we use CheXbert \cite{irvin2019chexpert} to annotate generated reports and compare them with ground truth labels across 14 categories, evaluating precision, recall, and F1 score.

\paragraph{Implementation Details.}
We adopt a ResNet-101 model pre-trained on ImageNet as the image encoder.
And the MedKLIP model \cite{wu2023medklipmedicalknowledgeenhanced} pre-trained on a widely used radiology report dataset is leveraged as the text encoder, initially aligning vision and language in the medical domain. 
The image size is set to 224, with feature channels of 768 for LRPs and GRPs and a fixed hyperbolic mapping channel of 512.
The coefficients $\alpha$ and $\beta$ are empirically set to 2 and 0.5, respectively. 
Optimization is performed using the AdamW optimizer with a weight decay of 0.05, an initial learning rate of $5e-5$, and a cosine learning rate schedule. Training runs for 6 epochs with a batch size of 18. All experiments are conducted on an NVIDIA A800 GPU (80GB) for about 10 hours using Python 3.10, PyTorch 2.4.0, and Ubuntu 22.04.

\textbf{Baseline.} We build a strong baseline by replacing CLIP with the MedKLIP foundation model from \cite{jin2024promptmrg} to incorporate medical domain knowledge. All our experiments are conducted on this baseline unless otherwise specified.

\subsection{Comparison with State-of-the-art Methods}
\paragraph{Quantitative Results.} 
To verify the effectiveness of our model, we compare its performance against various state-of-the-art (SOTA) models, including R2Gen \cite{chen2020generating}, M2TR \cite{nooralahzadeh2021progressive}, MKSG \cite{yang2022knowledge}, CliBert \cite{yan2022clinical}, CVT2Dis. \cite{nicolson2023improving}, M2KT \cite{yang2023radiology}, ME \cite{wang2023metransformer}, KiUT \cite{huang2023kiut}, DCL \cite{li2023dynamic}, RGRG \cite{tanida2023interactive}, UAR \cite{li2023unify},
HERGen \cite{wang2024hergenelevatingradiologyreport}, PromptMRG \cite{jin2024promptmrg}, and EKAGen \cite{bu2024instance}. 
Detailed comparison results on the MIMIC-CXR and IU X-Ray datasets are presented in Table \ref{tab1:env} and Table \ref{tab2:env}, respectively.
For the MIMIC-CXR dataset, as shown in Table \ref{tab1:env}, our proposed method achieves SOTA performance across all evaluation metrics, consistently outperforming the second-best approach by a large margin. Concretely, our method obtains the absolute improvements of 4.6\%, 6.0\%, 6.5\%, 6.3\%, 4.2\%, and 4.9\% over the recent work EKAGen on various NLG metrics. 
In terms of the CE metrics, our method outperforms the second-best EKAGen by 11.1\%, 13.0\%, and 9.3\% in Precision, Recall, and F1, respectively. 
For the IU X-Ray dataset, we follow PromptMRG \cite{jin2024promptmrg}
to evaluate on the entire dataset using pre-trained models from the MIMIC-CXR dataset. 
As illustrated in Table \ref{tab2:env}, our method achieves either the highest or runner-up performance across all NLG metrics and the highest performance on all CE metrics, surpassing the second-best model by 1.2\% and 6.3\% in the mean NLG and CE metrics, respectively.
Overall, our model consistently outperforms the second-best method, achieving the average improvement of 7.4\% and 2.9\% across all evaluation metrics on the MIMIC-CXR and IU X-Ray datasets, respectively.

\paragraph{Qualitative Results.} 
Figure \ref{fig3:env} presents two qualitative examples that highlight the superiority of our REVTAF model over both the baseline and the PromptMRG SOTA method. 
In Figure \ref{fig3:env}, text segments that completely align with the ground truth are highlighted in same colors, whereas non-corresponding segments are shown in black. 
As observed, our model effectively captures most of the key descriptions found in the ground truth. Specifically, it accurately identifies critical disease labels such as enlarged cardiomediastinum, pleural effusion, pneumothorax, as well as pertinent past medical history. 
In contrast, PromptMRG fails to accurately capture the past medical history and lung status, while the baseline generates several irrelevant details. 
Notably, our proposed method successfully eliminates these irrelevant sentences and produces more precise expressions, demonstrating the effectiveness of our approach.

\subsection{Ablation Study}
 \paragraph{Effectiveness of Each Component.} 
 We assess the effectiveness of each component in our method using the MIMIC-CXR dataset by incrementally incorporating them. The experimental results, summarized in Table \ref{tab3:env}, demonstrate that integrating the LRE module improves the baseline performance by 1.8\% and 0.9\% in average NLG and CE metrics, respectively, validating the advantage of adaptively retrieving the most relevant GRPs for report generation.
Similarly, the FVTAF module alone achieves gains of 1.7\% and 0.9\% over the baseline in mean NLG and CE metrics, respectively, emphasizing the effectiveness of fine-grained alignment and fusion.
When combining both the LRE and FVTAF modules, our REVTAF framework achieves significant improvements of 2.7\% and 2.9\% over the baseline in mean NLG and CE metrics, underscoring the critical role of these components in enhancing radiology report generation.
The proposed approach achieves an overall improvement of 2.8\% across all evaluation metrics compared to the baseline.
\paragraph{Comparison Results with multimodal LLMs.}
To validate the effectiveness of our method in radiology report generation, we compare it with mainstream multimodal LLMs, including GPT-4, GPT-4o, GPT-4o-mini, and GPT-4.5, using 16 randomly selected samples from the MIMIC-CXR test set (Table \ref{tab4:env}). For each sample, we measure report generation performance and inference time, averaging the results for comparison.
Our method consistently outperforms GPT-series multimodal LLMs across all metrics while being more efficient, making it better suited for clinical use. In contrast, GPT models, despite their extensive training, struggle with medical-specific knowledge, resulting in poorer performance and longer inference times. Among the GPT-4 series, smaller models (GPT-4o and GPT-4o-mini) underperform compared to larger ones (GPT-4 and GPT-4.5). While GPT-4o and GPT-4.5 slightly surpass GPT-4 in language metrics, they lag in diagnostic accuracy. Overall, GPT-4 performs best among these models but remains significantly inferior to our method.
\section{Conclusion}
We propose the Retrieval Enhanced Visual-Text Alignment and Fusion (REVTAF) framework for radiology report generation. REVTAF features a Learnable Retrieval Enhancer (LRE) to adaptively retrieve the most relevant GRPs, enhancing visual representations, especially for tail classes. Additionally, it employs a Fine-grained Visual-Text Alignment and Fusion (FVTAF) strategy, incorporating an FCC constraint for precise alignment and optimal transport-based cross-attention mechanism for improved fusion under weak supervision. Experiments show that REVTAF achieves favorable performance, delivering gains of 7.4\% and 2.9\% on the MIMIC-CXR and IU X-Ray datasets, significantly outperforming mainstream multimodal LLMs.

\section*{Acknowledgements}
This study was supported under the Key Research and development Project of Yunnan Province (Grant No. 202402AD080006).
It was also funded by the National Science Foundation of China (Grant No. 62201341).
{
    \small
    \bibliographystyle{ieeenat_fullname}
    \bibliography{main}
}

\end{document}


\maketitle

\section{Supplementary Ablation Studies}
We conducted additional ablation studies on the MIMIC-CXR test set, focusing on: 1) integrating the two proposed strategies into PromptMRG\cite{jin2024promptmrg} with CLIP, 2) analyzing key components, including hyperbolic distance and MPSA, and 3) exploring the effect of different batch size, and 4) the influence of hyperparameters.

\begin{figure}
  \centering
\includegraphics[width=0.47\textwidth]{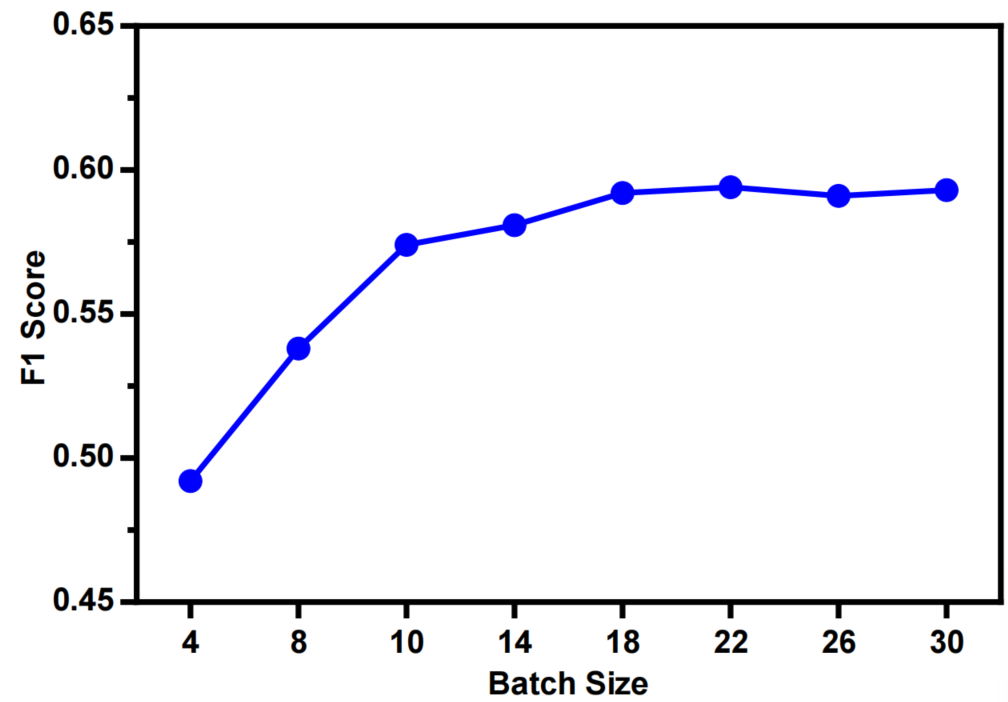}
   \caption{Effect of different batch size on model training performance.}
   \label{fig1:env}
\end{figure}

Table \ref{tab1:env} shows that, with the CLIP baseline, LRE alone improves average NLG and CE by 1.6\%, FVTAF alone by 1.3\%, and together achieve a 2.2\% gain. Table \ref{tab2:env} highlights that hyperbolic distance surpasses Euclidean distance and cosine similarity by 1.0\% and 0.6\%, respectively, demonstrating its strength in capturing hierarchical visual features. Table \ref{tab3:env} reveals that replacing traditional cross-attention with MPSA provides a further 0.5\% improvement, emphasizing the significance of both hyperbolic representations and MPSA.
Figure \ref{fig1:env} shows that performance declines with excessively small batches but plateaus as size increases. This is attributed to two factors: moderate batch sizes enhance input diversity through varied retrievals, while our FVTAF module’s multi-source alignment design filters noise, ensuring training robustness.

\begin{figure}
\centering
\includegraphics[width=0.47\textwidth]{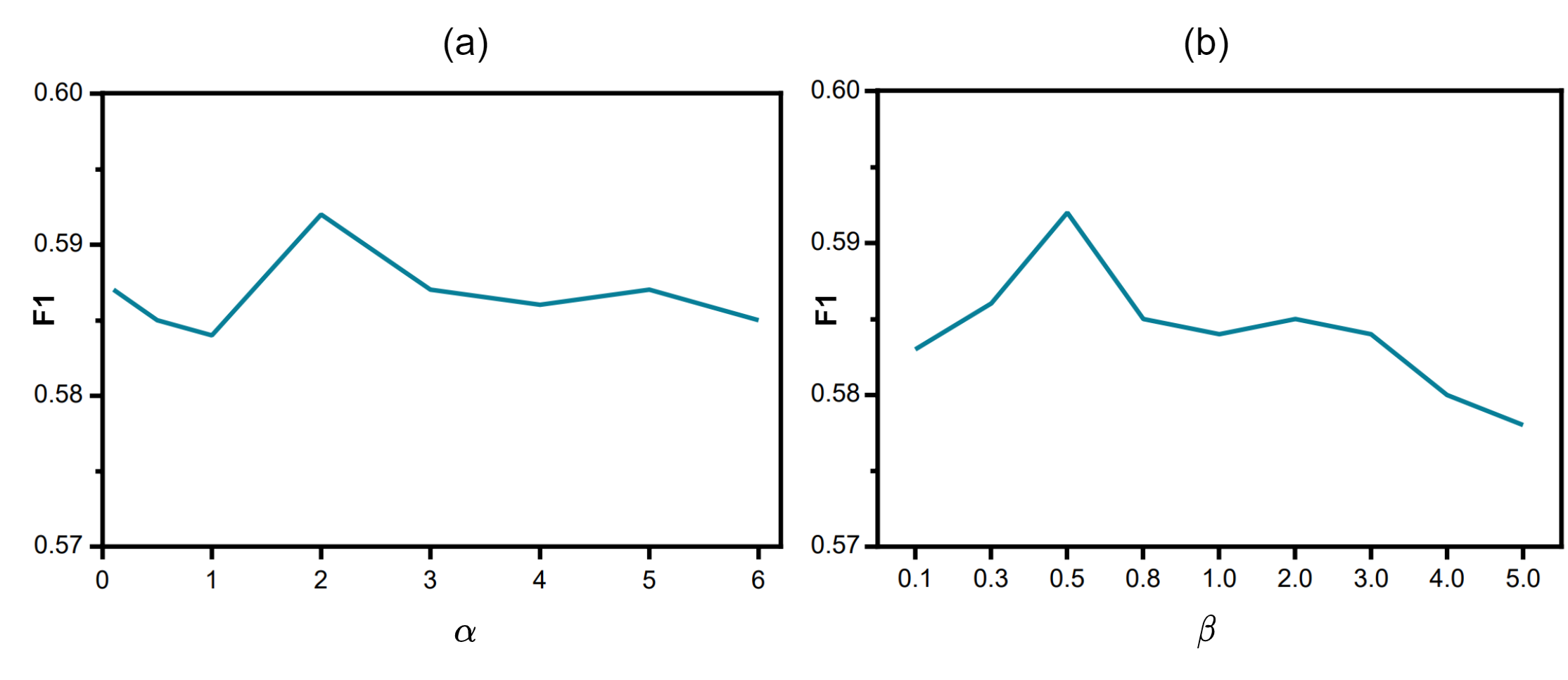}
\caption{Analysis of the hyperparameters $\alpha$ (in subfigure (a)) and $\beta$ (in subfigure (b)) with F1 scores on the MIMIC-CXR test set.}
\label{fig2:env}
\end{figure}

As shown in Figure \ref{fig2:env} (a) and (b),
the performance remains stable across a wide range, with F1 score fluctuations within 0.8\% for $\alpha$ and 1.4\% for $\beta$. 
Notably, the best F1 score of 0.592 is obtained when $\alpha = 2$ and $\beta=0.5$, which are the values we used in our experiments.
As illustrated in Figure \ref{fig2:env}(a), when $\alpha$ deviates from 2, either too small or too large, the selection of the global reference prompt is adversely affected, resulting in lower F1 performance. 
Similarly, Figure \ref{fig2:env}(b) shows that the optimal performance is reached at around $\beta=0.5$.
Values of $\beta$ that are too high or too low yield inferior outcomes, likely due to a multi-scale misalignment, which introduces noise and disrupts the report generation process.

\begin{table*}[ht]
    \centering
    \scalebox{0.8}{
    \begin{tabular}{l|cccccc|ccc|c}
        \toprule
        \multirow{2}{*}{Models}& \multicolumn{6}{c|}{NLG Metrics}  & \multicolumn{3}{c|}{CE Metrics }&\multirow{2}{*}{Avg}   \\
        \cline{ 2-10 }
         & BLEU-1 &BLEU-2 &BLEU-3 &BLEU-4 &METEOR &ROUGE&Precision &Recall &F1   \\ 
        \midrule
        Baseline (CLIP-based)&0.398 &0.239 &0.156	&0.112 &0.157 &0.268 &0.501	&0.509 &0.476&0.313 \\
        \midrule
          + LRE&0.419&0.252&0.171&0.119&0.178&0.281&0.532&0.521&0.490&0.329
   \\
         + FVTAF &0.407&0.251&0.169&0.119&0.163&0.275&0.524&0.533&\textcolor{magenta}{0.493}&0.326
     \\
        \midrule
         + LRE \& FVTAF  &\textcolor{magenta}{0.421}&\textcolor{magenta}{0.253}&	\textcolor{magenta}{0.175}&\textcolor{magenta}{0.121}&\textcolor{magenta}{0.185}&\textcolor{magenta}{0.286}&\textcolor{magenta}{0.535}&\textcolor{magenta}{0.537}&0.492&\textcolor{magenta}{0.335}
   \\ 
        \bottomrule
    \end{tabular}}
    \caption{Analysis on the effectiveness of each component within a CLIP-based foundation model on MIMIC-CXR test set.}
    \label{tab1:env}
\end{table*}

\begin{table*}[ht]
    \centering
    \scalebox{0.83}{
    \begin{tabular}{l|cccccc|ccc|c}
        \toprule
        \multirow{2}{*}{Methods} &\multicolumn{6}{c|}{NLG Metrics}  & \multicolumn{3}{c|}{CE Metrics }&\multirow{2}{*}{Avg}   \\
        \cline{ 2-10 }
                 & BLEU-1 &BLEU-2 &BLEU-3 &BLEU-4 &METEOR &ROUGE&Precision &Recall &F1   \\
        \midrule
        EDistance & 0.458&0.306&0.222&0.170&0.192&0.321&0.619&0.604&0.584&0.387
\\
        Cosine Similarity&0.461&0.315&0.232&0.181&0.197&0.333&0.618&0.603&0.583&0.391
 \\
        \midrule
        Hyperbolic        &\textcolor{magenta}{0.465} &\textcolor{magenta}{0.318} &\textcolor{magenta}{0.235} &\textcolor{magenta}{0.182} &\textcolor{magenta}{0.199} &\textcolor{magenta}{0.336} &\textcolor{magenta}{0.628}	&\textcolor{magenta}{0.613} &\textcolor{magenta}{0.592}&\textcolor{magenta}{0.397}  \\
        \bottomrule
    \end{tabular}}
    \caption{Comparison with different retrieval strategies on the MIMIC-CXR dataset.}
    \label{tab2:env}
\end{table*}

\begin{table*}[htb]
    \centering
    \scalebox{0.83}{
    \begin{tabular}{l|cccccc|ccc|c}
        \toprule
        \multirow{2}{*}{Model} &\multicolumn{6}{c|}{NLG Metrics}  & \multicolumn{3}{c|}{CE Metrics }&\multirow{2}{*}{Avg}   \\
        \cline{ 2-10 }
                 & BLEU-1 &BLEU-2 &BLEU-3 &BLEU-4 &METEOR &ROUGE&Precision &Recall &F1   \\
        \midrule
        Cross-attention &0.461&0.314&0.232&0.180&0.197&0.333&0.621&0.603&0.585&0.392
 \\
        MSPA (Ours)        &\textcolor{magenta}{0.465} &\textcolor{magenta}{0.318} &\textcolor{magenta}{0.235} &\textcolor{magenta}{0.182} &\textcolor{magenta}{0.199} &\textcolor{magenta}{0.336} &\textcolor{magenta}{0.628}	&\textcolor{magenta}{0.613} &\textcolor{magenta}{0.592}&\textcolor{magenta}{0.397}  \\
        \bottomrule
    \end{tabular}}
    \caption{Comparative evaluation of standard cross-attention and MPSA mechanisms on the MIMIC-CXR dataset.}
    \label{tab3:env}
\end{table*}

\begin{table}[htb]
    \centering
    \scalebox{0.9}{
    \begin{tabular}{l|l|cc}
        \toprule
        &Disease Classes & Samples &Distribution \\
        \midrule
        \multirow{7}{*}{Head Classes}&Enlarged Cardio&730&18.9\%	\\
        &Cardiomegaly&1271&32.9\% \\
        &Lung Opacity&1392&36.1\% \\
        &Edema&563&14.6\% \\
        &Atelectasis&841&21.8\% \\
        &Pleural Effusion&1056&27.4\% \\
        &Support Devices&1345&34.9\%  \\
        \midrule
        \multirow{7}{*}{Tail Classes}&Lung Lesion&199&5.2\%  \\
        &Consolidation&176&4.6\% \\
        &Pneumonia&165&4.3\% \\
        &Pneumothorax&75&1.9\% \\
        &Pleural Other&122&3.2\% \\
        &Fracture&148&3.8\% \\
        &No Finding&323&8.4\% \\
        \midrule
        Total& -&3858&- \\
        \bottomrule
    \end{tabular}}
    \caption{The number of samples and their distribution ratios across disease categories in the MIMIC-CXR test set.}
    \label{tab4:env}
\end{table}
\section{Tackling Data Imbalance}
Following the approaches \cite{chen2022cross,jin2024promptmrg}, we count the number of positive samples in the MIMIC-CXR test set and calculate the distribution of each disease, as detailed in Table \ref{tab4:env}.
The results reveal a pronounced long-tailed distribution, indicating the imbalance of disease classification.
For analytical clarity, we define diseases with a sample ratio exceeding 10\% as head classes, and those with lower proportions as tail classes. 
This categorization not only facilitates a more nuanced evaluation of our model's performance but also underscores the inherent challenges associated with imbalanced data in clinical imaging datasets.

\begin{figure}[htb]
\centering
\includegraphics[width=0.47\textwidth]{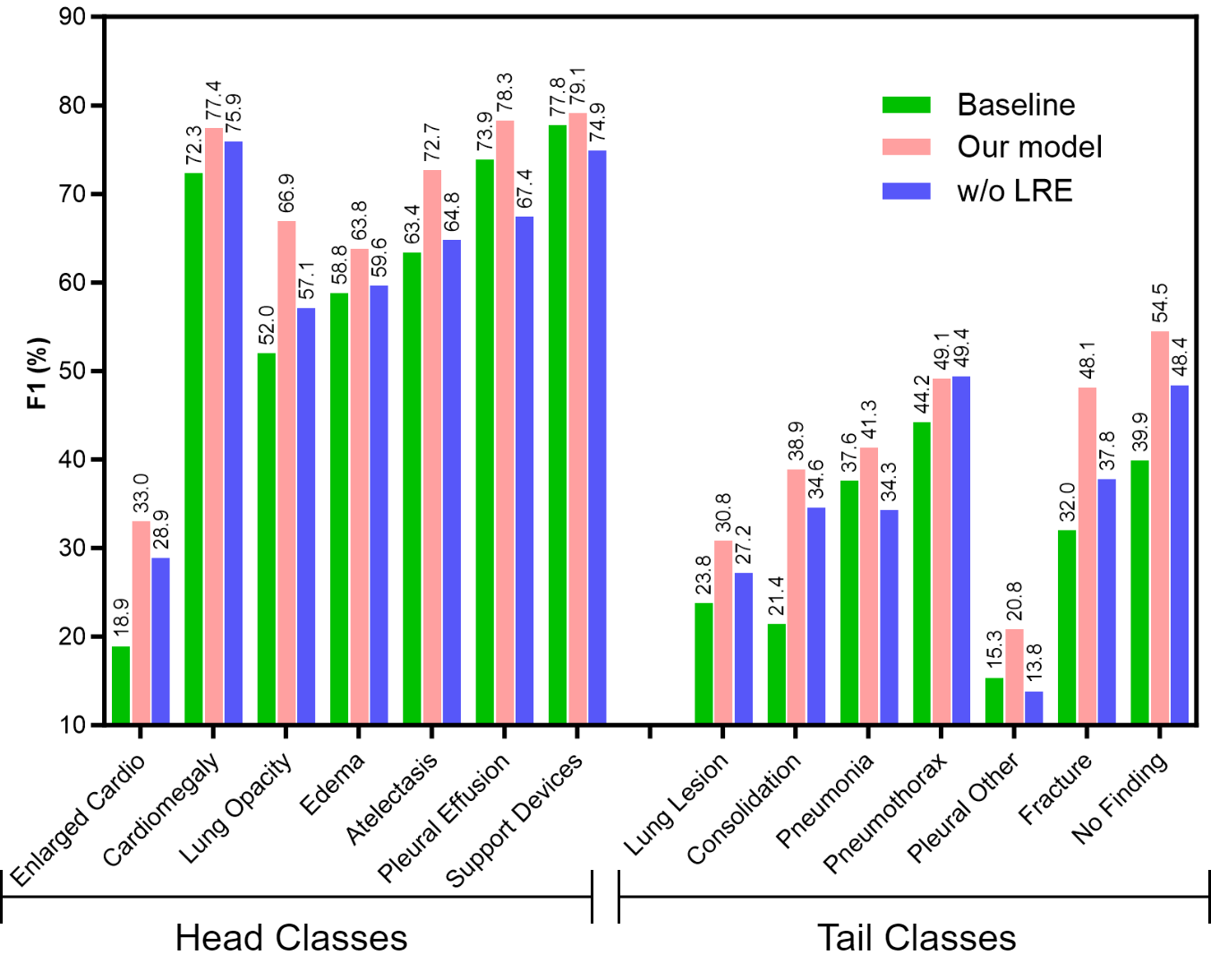}
\caption{Comparison of baseline and our method in addressing data imbalance, evaluated using F1 scores (\%). }
\label{fig3:env}
\end{figure}

\paragraph{Effectiveness of Addressing the Long-tailed Data Distribution.}
To evaluate the effectiveness of our method in addressing data imbalance, we categorized all diseases into head and tail groups based on sample sizes and compared the individual F1 scores between the baseline, our approach and w/o LRE module (Figure \ref{fig3:env}). 
Detailed head-tail grouping information is provided in the Table \ref{tab3:env}. As shown, our method consistently improves F1 scores across all disease classes, with a 9.9\% average increase for tail classes and a 7.7\% improvement for head classes compared to the baseline. Notably, the tail class ``Fracture" achieves a remarkable 16.1\% gain over the baseline. These results highlight the significant enhancements our framework brings to tail-class recognition while also delivering notable improvements for head classes.
Moreover, incorporating the LRE module consistently improves performance on both head and tail classes, with average gains of 5.43\% and 6.1\%, respectively, compared to our model, demonstrating its effectiveness in handling data imbalance. 

\section{Evaluation with More Advanced Metrics}
We adopt three advanced clinical evaluation metrics to comprehensively assess the effectiveness of our model, including 1/RadCliQ-v1\cite{Yu2022.08.30.22279318}, RadGraph\cite{Yu2022.08.30.22279318}, and BertScore\cite{zhang2020bertscoreevaluatingtextgeneration}).
Here we compare our model against several recent SOTA methods which are RGRG \cite{tanida2023interactive}, MedVersa\cite{zhou2025medversageneralistfoundationmodel}, and PromptMRG \cite{jin2024promptmrg}.
As shown in Table~\ref{tab5:env}, our approach achieves superior performance across all metrics, outperforming the strongest model (PromptMRG) by margins of 0.11, 0.10, and 0.08, respectively.
These results highlight the robustness and clinical applicability of our framework, demonstrating its capability to generate more accurate and semantically faithful reports in comparison to existing methods.
\begin{table}[htb]
    \centering
    \scalebox{0.8}{
    \begin{tabular}{l|ccc}
        \toprule
        Model  & 1/RadCliQ-v1 $\uparrow$ &RadGraph $\uparrow$ & BertScore $\uparrow$ \\
        \midrule
         RGRG & 0.76 & 0.17 & 0.35 \\
         MedVersa &  1.10 & 0.27 & 0.45 \\
         PromptMRG& 1.24 & 0.31 & 0.49 \\
        \midrule
        Ours        &\textcolor{magenta}{1.35} & \textcolor{magenta}{0.41} & \textcolor{magenta}{0.57}  \\
        \bottomrule
    \end{tabular}}
    \vspace{4.5ex} 
    \captionof{table}{Evaluation with advanced clinic scores on the MIMIC-CXR test set.}
    \label{tab5:env}
\end{table}

\section{Supplementary Instruction for Evaluation on GPT-Series Multi-Modal LLMs}
We use a consistent prompt for GPT-series multi-modal LLMs: ``[You are helpful assistant of a radiologist. Your task is help the radiologist to draft the professional radiology report.] + [ Radiology image ] + [The image above is an X-ray a patients. Write a professional report on it. Answer in one paragraph, and only include the finding part.]''.
However, the generated output often contains extraneous information, making the evaluation unfair. 
To ensure consistency, we performed a two-step data cleaning process:
(1) Extract only the ``Findings" section from the reports and consolidate it into an individual report.
(2) Remove numbers, line breaks, and other unnecessary elements, limiting the text length to 200 characters.
The final evaluation results, as presented in the main paper, are reported using the same evaluation criteria as previous studies \cite{jin2024promptmrg,bu2024instance}.

{
    \small
    \bibliographystyle{ieeenat_fullname}
    \bibliography{main}
}